\documentclass[journal]{IEEEtran}

\usepackage{acronym}
\usepackage{algorithm}
\usepackage[noend]{algpseudocode}
\usepackage{amsfonts,amsmath,amssymb}
\usepackage{array}
\usepackage{bm}
\usepackage{booktabs}
\usepackage{capt-of}
\usepackage{cite}
\usepackage{color}
\usepackage{comment}
\usepackage{dsfont}
\usepackage{esvect}
\usepackage{graphicx}
\usepackage[hidelinks]{hyperref}
\usepackage{lipsum}
\usepackage{multirow}
\usepackage{nicefrac}
\usepackage[Gray,amssymb]{SIunits}
\usepackage[caption=false,font=normalsize,labelfont=sf,textfont=sf]{subfig}
\usepackage{stfloats}
\usepackage{textcomp}
\usepackage{url}
\usepackage{verbatim}
\usepackage[table]{xcolor}

\definecolor{lightgray}{HTML}{FFFFFF}
\definecolor{mediumgray}{HTML}{E0E0E0}
\definecolor{darkgray}{HTML}{B0B0B0}
\newcommand{\lc}[1]{\cellcolor{lightgray}#1}
\newcommand{\mc}[1]{\cellcolor{mediumgray}#1}
\newcommand{\dc}[1]{\cellcolor{darkgray}#1}

\acrodef{AC}{alternating current}
\acrodef{ACOPF}[AC-OPF]{AC optimal power flow}
\acrodef{CDHR}{coordinate directions hit-and-run}
\acrodef{CLT}{central limit theorem}
\acrodef{DC}{direct current}
\acrodef{EHV}{extra high voltage}
\acrodef{FC}{fully connected}
\acrodef{FCNN}{fully connected neural network}
\acrodef{GNN}{graph neural network}
\acrodef{GPU}{graphics processing unit}
\acrodef{HV}{high voltage}
\acrodef{IPM}{interior point method}
\acrodef{KCL}{Kirchhoff current law}
\acrodef{KDE}{kernel density estimate}
\acrodef{KKT}{Karush-Kuhn-Tucker}
\acrodef{KPCA}{kernel principal component analysis}
\acrodef{LRI}{long-range interaction}
\acrodef{ML}{machine learning}
\acrodef{MP}{message passing}
\acrodef{MPNN}{message passing GNN}
\acrodef{MLP}{multi-layer perceptron}
\acrodef{NLP}{non-linear programming}
\acrodef{NN}{neural network}
\acrodef{OPF}{optimal power flow}
\acrodef{OOD}{out of distribution}
\acrodef{QC}{quadratic convex}
\acrodef{TSO}{transmission system operator}
\acrodef{URS}{uniform random sampling}
\acrodef{VOLL}{value of lost load}

\begin{document}

\title{A Principled Framework to Evaluate Quality of\\AC-OPF Datasets for Machine Learning:\\Benchmarking a Novel, Scalable Generation Method}
\author{
    Matteo~Baù, Luca~Perbellini and Samuele~Grillo,~\IEEEmembership{Senior Member,~IEEE}
    \thanks{This work has been partly financed by the Research Fund for the Italian Electrical System under the Three-Year Research Plan 2022--2024 (DM MITE n. 337, 15.09.2022), in compliance with the Decree of April 16th, 2018, and by the EU funds Next-GenerationEU (Piano Nazionale di Ripresa e Resilienza (PNRR) – Missione 4 Componente 2, Linea d'Investimento 3.3) --- D.M. 352 09/04/2022.}
    \thanks{M.~Baù is with Ricerca sul Sistema Energetico S.p.A., 20134 Milan, Italy (e-mail: matteo.bau@rse-web.it).} 
    \thanks{L.~Perbellini and S.~Grillo are with Politecnico di Milano, DEIB, 20133 Milan, Italy (e-mail: {name}.{surname}@polimi.it).}
    \thanks{This paper is an extended version of \cite{GEN-Ours}.}
}

\maketitle

\begin{abstract}
    Several methods have been proposed in the literature to improve the quality of \ac{ACOPF} datasets used in \ac{ML} models. Yet, scalability to large power systems remains unaddressed and comparing generation approaches is still hindered by the absence of widely accepted metrics quantifying \ac{ACOPF} dataset quality. In this work, we tackle both these limitations. We provide a simple heuristic that samples load setpoints uniformly in total load active power, rather than maximizing volume coverage, and solves an \ac{ACOPF} formulation with load slack variables to improve convergence. For quality assessment, we formulate a multi-criteria framework based on three metrics, measuring variability in the marginal distributions of \ac{ACOPF} primal variables, diversity in constraint activation patterns among \ac{ACOPF} instances and activation frequency of variable bounds. By comparing four open-source methods based on these metrics, we show that our heuristic consistently outperforms uniform random sampling, whether independent or constrained to a convex polytope, scoring as best in terms of balance between dataset quality and scalability. 

\end{abstract}


\section{Introduction}\label{sec:intro}

\IEEEPARstart{T}{he} \ac{ACOPF} is a central problem in power grids optimization, planning and market clearing. Despite being first introduced over 60 years ago \cite{OPF-Carpentier(1962)}, the inherent non-convexities and nonlinearities in the AC power flow equations continue to hinder its practical application to real-scale power systems \cite{OPF-Lavaei/Duality}. As a result, the industry has traditionally relied on linearized approximations, such as the DC-OPF. However, these simplified models can lead to increased system operation costs \cite{OPF-Cain/History}, prompting the development of a wide range of alternative OPF formulations, such as convex relaxations, to address these limitations \cite{OPF-Capitanescu/Review}.

In recent years, \ac{ML} has emerged as a promising alternative to solve the \ac{ACOPF} problem, offering the potential to shift the computational burden to the offline training phase. Various \ac{ML}-based approaches have been proposed in the literature, including methods to warm-start traditional solvers \cite{NN-Baker/Warm, PINN-Dong/PGSim, NN-Deihim/WarmStart}, reduce problem dimensionality through active constraints prediction \cite{NN-Falconer/Leveraging, NN-Fouad/ActiveSet} or directly approximate the \ac{ACOPF} solution \cite{NN-Zamzam/Learning, PINN-Fioretto/Dual, NN-Owerko/GNN, NN-Falconer/Leveraging, PINN-Donti/DC3, PINN-Nellikkath/Physics, NN-Singh/Sensitivity, PINN-Pan/DeepOPF, PINN-Liu/GNNDual, PINN-Owerko/GNN, PINN-Garcia/TypedGNN, PINN-Piloto/CANOS, PINN-Pareek/Bayesian, PINN-Park/PrimalDual, NN-Rahman2/Augmented, PINN-Yang/GuidedGNN}.
Among these, direct approximation offers the greatest potential for computational speedup and has drawn significant research interest, partly due to its inherent modeling complexity. Specifically, it requires careful balancing between optimality and constraint feasibility during \ac{ML} training \cite{PINN-Park/PrimalDual}. This is typically tackled by learning the mapping from load input setpoints to optimal primal variables, penalizing constraint violations in the loss function, either in supervised \cite{PINN-Nellikkath/Physics, PINN-Piloto/CANOS} or self-supervised \cite{PINN-Liu/GNNDual, PINN-Owerko/GNN, PINN-Garcia/TypedGNN, PINN-Park/PrimalDual} learning settings.

Regardless of the architecture, \ac{ML} model performance for the \ac{ACOPF} strongly depends on the quality of the datasets used for training and testing \cite{GEN-Jones/OPFLearn, GEN-Ignasi/Rambo}. In the majority of studies, such datasets are synthetically generated by sampling uniformly at random each load's active and reactive power within a specified range, as shown in Table \ref{tab1:variation_range}, around the respective nominal values. In \cite{NN-Zamzam/Learning, PINN-Donti/DC3} this base method is modified by using a truncated Gaussian distribution, while \cite{PINN-Nellikkath/Physics} employs Latin hypercube sampling. Differently, the authors of \cite{NN-Rahman2/Augmented, PINN-Yang/GuidedGNN} obtain \ac{ACOPF} instances based on aggregated historical load data to better reflect realistic patterns.

\begin{table}[ht]
    \caption{Range of variation used in reviewed works}\label{tab1:variation_range}
    \centering
    \begin{tabular}{cc}
        \toprule
        \textbf{Range [\%]} & \textbf{Reference}\\
        \midrule
        $\pm 10$ & \cite{PINN-Dong/PGSim, NN-Owerko/GNN, PINN-Pan/DeepOPF, PINN-Owerko/GNN, PINN-Liu/GNNDual} \\
        $\pm 20$ & \cite{NN-Baker/Warm, PINN-Fioretto/Dual, NN-Falconer/Leveraging, PINN-Nellikkath/Physics, NN-Singh/Sensitivity, PINN-Garcia/TypedGNN, PINN-Piloto/CANOS, PINN-Pareek/Bayesian} \\
        $\pm 70$ & \cite{NN-Zamzam/Learning, PINN-Donti/DC3} \\
        \bottomrule
    \end{tabular}
\end{table}

Despite their widespread use, these methods tend to generate \ac{ACOPF} datasets with notable limitations that may impair the generalization capabilities of \ac{ML} models, reducing confidence in their performance \cite{GEN-Jones/OPFLearn, GEN-Ignasi/Rambo}. In particular, recent studies have shown that the resulting \ac{ACOPF} instances cover only a limited portion of the feasible space \cite{GEN-Jones/OPFLearn} and fail to adequately capture the operational limits of power systems \cite{GEN-Ignasi/Rambo}.

To address these issues, multiple methodologies have been proposed in the literature. \textit{Historical} sampling is currently employed only in \cite{GEN-Gillioz/Entsoe}, due in part to the scarcity of open-source operational power system data and in part to the inherent risk of training \ac{ML} models on outdated patterns \cite{GEN-Bugaje/Balancing}. \textit{Generic} sampling, on the other hand, is more utilized and aims to maximize coverage of the \ac{ACOPF} feasible space, though possibly at the expense of realism and diversity in the generated instances \cite{GEN-Bugaje/Balancing}.

The authors of OPFLearn \cite{GEN-Jones/OPFLearn} achieve high coverage by sampling uniformly within the load input space, defined as a convex polytope that is iteratively reduced using infeasibility certificates, following the procedure of \cite{GEN-Venzke/Efficient}. This is extended in \cite{GEN-Ours} by enforcing uniformity in total load active power, similarly to \cite{NN-Deihim/WarmStart}, and by improving convergence based on a direct walk approach. In \cite{GEN-Bugaje/Split}, power flow samples are generated sequentially as centered within the largest, empty subregion of the feasible space. RAMBO \cite{GEN-Ignasi/Rambo}, instead, casts the dataset generation as a bi-level optimization problem, finding new \ac{ACOPF} instances that are maximally distant from previous ones in terms of load active power, generator active power and bus voltage magnitude. Lastly, in \cite{GEN-Nellikkath/Enriching, GEN-Hu/Worth, GEN-Zhang/CSS} \acp{NN} are integrated into the sampling process to identify critical instances that enhance dataset coverage and informativeness. 

Each methodology evaluates the quality of the generated \ac{ACOPF} datasets relative to the uniform sampling baseline approach, typically using custom metrics and, in some cases, empirical testing with \ac{ML} models. To this end, OPFLearn \cite{GEN-Jones/OPFLearn} computes the number of unique combinations of active inequality constraints. Similarly, RAMBO \cite{GEN-Ignasi/Rambo} and \cite{GEN-Ours} count how many constraints become active at least once in the dataset, differentiating across \ac{ACOPF} primal variables. More general metrics are adopted in other contributions: \cite{GEN-Bugaje/Split} measures coverage by computing the volume of the convex hull enclosing the samples, while \cite{GEN-Zhang/CSS} employs the Simpson-diversity index, which is applied after clustering the \ac{ACOPF} instances based on \ac{KPCA} and k-means. 

For empirical validation, the typical approach is to train two \ac{ML} models separately, one on the proposed dataset and the other on the baseline one, comparing their performance. In \cite{GEN-Hu/Worth} the \ac{ML} models are tested on a common split containing \ac{OOD} instances, whereas in \cite{GEN-Jones/OPFLearn} and \cite{GEN-Zhang/CSS} the model trained on a dataset is tested on the other.

Overall, no advanced methodology for generating \ac{ACOPF} datasets has emerged as clear choice in the literature, and uniform random sampling remains widely utilized, notwithstanding its known limitations. To the best of our knowledge, existing \textit{generic} sampling approaches have not been evaluated on power systems exceeding 118–300 buses, leaving the question of scalability largely unaddressed. Furthermore, direct comparison across different methods is hindered by the lack of standard, widely accepted metrics for assessing the quality of \ac{ACOPF} datasets in the context of \ac{ML} models training. In particular, it remains unclear how different \textit{generic} sampling strategies perform in terms of diversity in the AC-OPF instances \cite{GEN-Bugaje/Balancing}.

Motivated by these research gaps, this paper aims to advance the state-of-the-art in \ac{ACOPF} dataset generation and quality analysis. Specifically, it extends our previous methodology introduced in \cite{GEN-Ours} by significantly improving its scalability, while also addressing the broader need for standardized dataset evaluation and comparison. Overall, the contributions of our paper are threefold:
\begin{enumerate}
    \item We propose a simple yet effective \textit{generic} sampling heuristic to generate high-quality \ac{ACOPF} datasets, increasing diversity through uniform sampling in total load active power as in \cite{GEN-Ours} and improving scalability by avoiding altogether the need to reduce the input load convex polytope. The method is validated on power system test cases with over 4000 buses.
    \item We introduce three informative, interpretable and complementary metrics to assess the quality of \ac{ACOPF} datasets used in \ac{ML} applications.
    \item We perform a comprehensive and robust comparison among multiple open-source methodologies for generating AC-OPF datasets, namely the baseline uniform random sampling, OPFLearn \cite{GEN-Jones/OPFLearn}, RAMBO \cite{GEN-Ignasi/Rambo} and our approach.
\end{enumerate}

To support reproducibility and adoption, we release the implementation, alongside some support documentation, as open-source at the GitHub repository \textit{HEDGeOPF.jl} \cite{PKG-HEDGeOPF}.

The remainder of the paper is organized as follows. Section \ref{sec:problem} characterizes a core issue that limits diversity in \textit{generic} sampling methods for the AC-OPF. Section \ref{sec:method} details the proposed methodology, addressing this limitation. Section \ref{sec:metrics} presents the metrics that are used in Section \ref{sec:comparison} to compare the quality of AC-OPF datasets generated with selected open-source methods. Finally, Section \ref{sec:conclusions} concludes this work, summarizing the main results and contributions.

\section{Problem statement}\label{sec:problem}

As highlighted in \cite{GEN-Bugaje/Balancing}, coverage maximization with \textit{generic} sampling may conflict with diversity, where the latter defines how much information the dataset samples hold. This trade-off is inherently dependent on the properties of the application domain. In the case of the \ac{ACOPF} problem, it is well-known that the optimal solution is heavily influenced by the total active power loading, primarily through the economic dispatch subproblem \cite{OPF-Cain/History}. Consequently, systematically varying the total active loading is expected to increase diversity in the resulting \ac{ACOPF} dataset, especially in terms of generator active power setpoints.

Under standard random sampling, however, the variation in system-wide consumption is limited, shrinking alongside system size. According to the \ac{CLT}, the distribution of the total active power load across \ac{ACOPF} instances generated via independent, random sampling is approximately Gaussian, centered at the sum of the means of the $|\mathcal{D}|$ individual load distributions and characterized by a coefficient of variation scaling as $1/\sqrt{|\mathcal{D}|}$. This holds true for independent distributions with finite variance and as long as no single load dominates the sum (see \cite{VOL-Vershynin/Probability}, Ch. 2). A similar phenomenon arises even when sampling uniformly within the convex polytope defined by the load space, due to the concentration of measure effect. In high-dimensional spaces, this effect implies that most of the volume of the convex body concentrates sharply near its center and, thus, uniform sampling yields total load active power values that cluster tightly around a central value (see \cite{VOL-Vershynin/Probability}, Ch. 5).

Altogether, this phenomenon stands for a primary factor explaining the limited diversity of AC-OPF datasets generated via uniform random sampling, regardless of whether load variables are sampled independently or jointly (i.e., within a convex polytope). In particular, this suggests that:
\begin{itemize}
    \item Methods focusing solely on load space coverage, such as OPFLearn \cite{GEN-Jones/OPFLearn} and \cite{GEN-Bugaje/Split}, are not expected to outperform the baseline in terms of dataset diversity. This is confirmed empirically in Section \ref{sec:comparison} for OPFLearn \cite{GEN-Jones/OPFLearn}.
    \item Explicit control over total active power load should be incorporated into the sampling process to ensure sufficient variability in this dimension and, consequently, higher diversity in the generated \ac{ACOPF} datasets.
\end{itemize}

These observations motivate the development of a sampling approach that directly enforces variability in total load active power, forming the basis of the methodology proposed in the Section \ref{sec:method}.

\section{Methodology}\label{sec:method}

This section presents the methodology developed for generating high-quality \ac{ACOPF} datasets, addressing both the lack of diversity discussed in Section \ref{sec:problem} and the need for computational scalability. The heuristic for sampling in the load space is introduced first. Then, the \ac{ACOPF} formulation adopted in this work and the instance generation process are detailed.

\subsection{Load space sampling}\label{subsec:sampling}

Active and reactive load setpoints are sampled from a convex polytope, following a similar approach to \cite{GEN-Venzke/Efficient, GEN-Jones/OPFLearn}. This consists in a H-polytope of the form $\bm{A}\bm{x} \leq \bm{b}$ with $\bm{x}\in\mathbb{R}^{2|\mathcal{D}|}$, bounded by limits on active power, reactive power and power factor of each load $d$ and on total load active power. In particular, the bounds on load active (reactive) power are centered around the nominal value $\bar{p}_d$ ($\bar{q}_d$) and symmetrically scaled by a fixed percentage $\delta_p$ ($\delta_q$), analogous to the range used in standard independent random sampling: 

\begin{subequations}\label{eq-3A:load_limits}
\begin{align}
    \left(1 - \frac{\delta_p}{100}\right) \leq \frac{p_d}{\bar{p}_d} \leq \left(1 + \frac{\delta_p}{100}\right) \quad &\forall d \in \mathcal{D} \label{eq-3A:pd_limit} \\
    \left(1 - \frac{\delta_q}{100}\right) \leq \frac{q_d}{\bar{q}_d} \leq \left(1 + \frac{\delta_q}{100}\right) \quad &\forall d \in \mathcal{D} \label{eq-3A:qd_limits}
\end{align}
\end{subequations}

The power factor of a given load is constrained between minimum and maximum values dependent on the nominal one, with the phase angle $\bar{\theta}_d$ defined by $\arctan(\bar{q}_d/\bar{p}_d)$:

\begin{subequations}\label{eq-3A:pf_limits}
\begin{align}
    \cos\left(\theta^{\rm min}_d\right) &= \max\left(\cos\left(\bar{\theta}_d\right)-\delta_{\text{pf}}, \alpha_{\rm min}\right) \\
    \cos\left(\theta^{\rm max}_d\right) &= \max\left(\cos\left(\bar{\theta}_d\right), \alpha_{\rm max}\right)
\end{align}
\end{subequations}
where $\alpha_{\rm min}$ and $\alpha_{\rm max}$ are user-defined, nonnegative power factor values preventing very large and very small, unrealistic reactive power demands, respectively. The nonnegative parameter $\delta_{\text{pf}}$ reduces the nominal power factor, thereby enlarging the range of reactive power variation. Like in \cite{GEN-Jones/OPFLearn, GEN-Ours}, the bounds in \eqref{eq-3A:pf_limits} are then reformulated as limits on the ratio between reactive and active power, coupling the two variables and,  differently from \cite{GEN-Jones/OPFLearn}, preserving the nominal lagging or leading behavior. The actual lower and upper bounds in \eqref{eq-3A:qp_limits} depend on the sign of the ratio for a given load and are labeled as $r_d^{\rm min}$ and $r_d^{\rm max}$.

\begin{equation}\label{eq-3A:qp_limits}
    \tan\left(\theta^{\rm max}_d\right) \leq \operatorname{sign}\left(\frac{\bar{q}_d}{\bar{p}_d}\right)\frac{q_d}{p_d} \leq \tan\left(\theta^{\rm min}_d\right) \quad \forall d \in \mathcal{D} 
\end{equation}

To increase variance in total active power load across input setpoints, sampling is not performed uniformly in the convex polytope defined by \eqref{eq-3A:load_limits} and \eqref{eq-3A:qp_limits}. Instead, at first $K$ total active power load samples $\bm{p}^{\text{tot}}=\left[p_1^{\text{tot}}, \dots, p_K^{\text{tot}} \right]^\intercal \in \mathbb{R}^K$ are drawn from a uniform distribution defined over the support:

\begin{align}\label{eq-3A:distribution}
    \bm{p}^{\text{tot}} \sim \mathcal{U}\Bigl(
        \max&\left({\bm{p}_d^{\rm min}}^\intercal\bm{1}_{|\mathcal{D}|},  \;{\bm{p}_g^{\rm min}}^\intercal\bm{1}_{|\mathcal{G}|}\right), \\ \min&\left({\bm{p}_d^{\rm max}}^\intercal\bm{1}_{|\mathcal{D}|}, \;{\bm{p}_g^{\rm max}}^\intercal\bm{1}_{|\mathcal{G}|}\right) \notag
    \Bigr)
\end{align}
where $\bm{p}_g^{\rm min}$ and $\bm{p}_g^{\rm max}$ are vectors of lower and upper limits of generator active power, with $\mathcal{G}$ being the generator set, while $\bm{p}_d^{\rm min}$ and $\bm{p}_d^{\rm max}$ are the load active power bounds given in \eqref{eq-3A:pd_limit}. The convex polytope defined by \eqref{eq-3A:load_limits} and \eqref{eq-3A:qp_limits} is then sliced by adding a new inequality on total load active power, constraining its variation in a neighborhood $\epsilon$ of a given sampled value $p_k^{\text{tot}}$:

\begin{equation}\label{eq-3A:pdtot_limit}
    p_k^{\text{tot}} - \epsilon \leq \sum_{d=1}^{\mathcal{D}} p_d \leq p_k^{\text{tot}} + \epsilon
\end{equation}

Combining constraints \eqref{eq-3A:load_limits}, \eqref{eq-3A:qp_limits} and \eqref{eq-3A:pdtot_limit} yields the final matrix representation \eqref{eq-3A:polytope} of the convex polytope slice that is sampled uniformly to generate active and reactive power load setpoints $\bm{x}=\left[\bm{p}_d^\intercal, \bm{q}_d^\intercal\right]^\intercal$. The \ac{CDHR} random walk is used for sampling, since it entails a low computational cost per sample without compromising on mixing time in practice \cite{VOL-Chalkis/Volesti}.

\begin{equation}\label{eq-3A:polytope}
    \bm{A} := \left[
        \begin{array}{@{}rr@{}}
        \bm{\mathrm{I}}_{|\mathcal{D}| \times |\mathcal{D}|} & \bm{0}_{|\mathcal{D}| \times |\mathcal{D}|} \\
        -\bm{\mathrm{I}}_{|\mathcal{D}| \times |\mathcal{D}|} & \bm{0}_{|\mathcal{D}| \times |\mathcal{D}|} \\
        \bm{0}_{|\mathcal{D}| \times |\mathcal{D}|} & \bm{\mathrm{I}}_{|\mathcal{D}| \times |\mathcal{D}|} \\
        \bm{0}_{|\mathcal{D}| \times |\mathcal{D}|} & -\bm{\mathrm{I}}_{|\mathcal{D}| \times |\mathcal{D}|} \\
        -\text{diag}\left(\bm{r}_d^{\rm max}\right) & \bm{\mathrm{I}}_{|\mathcal{D}| \times |\mathcal{D}|} \\
        \text{diag}\left(\bm{r}_d^{\rm min}\right) & -\bm{\mathrm{I}}_{|\mathcal{D}| \times |\mathcal{D}|} \\
        \bm{1}_{1 \times |\mathcal{D}|} & \bm{0}_{1 \times |\mathcal{D}|} \\
        -\bm{1}_{1 \times |\mathcal{D}|} & \bm{0}_{1 \times |\mathcal{D}|}
        \end{array}
    \right], \quad
    \bm{b} := \left[
        \begin{array}{@{}c@{}}
        \bm{p}_d^{\rm max} \\
        -\bm{p}_d^{\rm min} \\
        \bm{q}_d^{\rm max} \\
        -\bm{q}_d^{\rm min} \\
        \bm{0}_{|\mathcal{D}|} \\
        \bm{0}_{|\mathcal{D}|} \\
        p_k^{\text{tot}} + \epsilon \\
        \epsilon - p_k^{\text{tot}} 
        \end{array}
    \right]
\end{equation}

It must be highlighted that uniform sampling from a polytope slice leads to clustering of samples on the total load active power limit closest to the polytope centroid, due to the concentration of measure effect \cite{VOL-Vershynin/Probability}. Therefore, accurately approximating a uniform distribution in total active power load requires applying a small value of $\epsilon$, which is set to 0.1\% of the support range specified in \eqref{eq-3A:distribution} in this work.

Overall, by generating load samples for different values of $p_k^{\text{tot}}$, this approach ensures control over system-wide loading conditions while preserving multiple individual and coupling constraints. It is worth noting, however, that sampling based on total active power load is not new per se and has already been applied al least once, specifically in \cite{NN-Deihim/WarmStart}, to generate \ac{ACOPF} datasets for ML. In that work, individual active load values are drawn from a Dirichlet distribution that satisfies a fixed total active power value, but cannot account for negative loads (i.e., renewable generation) or enforce individual active power constraints. This limits its applicability in realistic settings and its usage in dataset comparison under consistent assumptions regarding the load space boundaries. 

\subsection{AC-OPF instance generation}\label{subsec:formulation}

Depending on the specific power system test case, the convex load sampling space defined by \eqref{eq-3A:polytope} may be mostly unfeasible with respect to the \ac{ACOPF} problem \cite{GEN-Venzke/Efficient, GEN-Jones/OPFLearn}. This issue worsens with system size \cite{GEN-Venzke/Efficient, GEN-Ours}, potentially leading to low convergence rates when generating datasets of feasible \ac{ACOPF} instances and significantly increasing the overall computational cost. Prior works \cite{GEN-Venzke/Efficient, GEN-Jones/OPFLearn} and \cite{GEN-Ours} address this limitation by iteratively trimming the convex polytope with hyperplanes, which represent infeasibility certificates generated based on a \ac{QC} relaxation of the \ac{ACOPF}. In addition, \cite{GEN-Venzke/Efficient} and \cite{GEN-Ours} solve a direct-walk problem to find the closest feasible \ac{ACOPF} solution if the input setpoint sampled from the reduced polytope is still unfeasible.

In this work, we completely bypass this procedure by sampling directly from the unclassified load space \eqref{eq-3A:polytope}, using the approach described in Section \ref{subsec:sampling}, and generating dataset instances based on an \ac{ACOPF} formulation with complex load power slack variables $\bm{s}_d^{\text{up}}$ and $\bm{s}_d^{\text{dw}}$:

\begin{subequations}\label{eq-3B:opf}
    \begin{alignat}{2}
        \min_{\bm{v}_n,\bm{s}_g,\bm{s}_d} & \Bigl(\bm{p}_g^\intercal\bm{C}_{g2}\bm{p}_g + \bm{c}_{g1}^\intercal\bm{p}_g + \bm{c}_{g0}^\intercal\bm{1_{|\mathcal{G}|}} \hspace{-15em} & \label{eq-3B:objective} \\
        & +\bm{c}_d^\intercal\left(\bm{p}_d^{\text{up}}+ \bm{q}_d^{\text{up}} + \bm{p}_d^{\text{dw}} + \bm{q}_d^{\text{dw}}\right)\Bigr) \hspace{-20em} & \notag \\
        \text{s.t.}\hspace{0.65em} & \eqref{eq-3A:load_limits}, \eqref{eq-3A:qp_limits}, \eqref{eq-3A:pdtot_limit} & \notag  \\
        & \bm{W}_g\bm{s}_g - \bm{W}_d\bm{s}_d = \bm{v}_n\bm{Y}_b^*\bm{v}_n^* \hspace{-5em} & \label{eq-3B:kcl}\\
        & s_{ij} = Y_{ij}^*v_i^*v_i^* - Y_{ij}^*v_i^*v_j^* \quad & \forall (i,j) \in \mathcal{E} \cup \mathcal{E}^R \label{eq-3B:power_flow} \\
        & s_d = \hat{s}_d + s_d^{\text{dw}} - s_d^{\text{up}} & \forall d \in \mathcal{D} \label{eq-3B:slack} \\
        & v_n^{\rm min} \leq |v_n| \leq v_n^{\rm max} & \forall n \in \mathcal{N} \label{eq-3B:vm_limits} \\
        & p_g^{\rm min} \leq p_g \leq p_g^{\rm max} & \forall g \in \mathcal{G} \label{eq-3B:pg_limits} \\
        & q_g^{\rm min} \leq q_g \leq q_g^{\rm max} & \forall g \in \mathcal{G} \label{eq-3B:qg_limits} \\
        & |s_{ij}| \leq s_{ij}^{\rm max} & \forall (i,j) \in \mathcal{E} \cup \mathcal{E}^R \label{eq-3B:sb_limits} \\
        & \theta_{ij}^{\rm min} \leq \theta_i - \theta_j \leq \theta_{ij}^{\rm max} & \forall (i,j) \in \mathcal{E} \label{eq-3B:theta_limits} \\
        & p_d^{\text{up}}, \ p_d^{\text{dw}}, \ q_d^{\text{up}}, \ q_d^{\text{dw}} \geq 0 & \forall d \in \mathcal{D}
    \end{alignat}
\end{subequations}
where $\mathcal{N}$ and $\mathcal{E}$ are the sets of nodes and branches of the given power system, with $\mathcal{E}^R$ being the set with reverse orientation.
$\bm{Y}_b$ and $\bm{Y}$ are the complex bus and branch admittance matrices, respectively. The binary connection matrices $\bm{W}$ map each element of a given type to the respective node(s) of connection. For example, the element $(n,i)$ of the generator connection matrix $\bm{W}_g \in \{0,1\}^{|\mathcal{N}| \times |\mathcal{G}|}$ is equal to 1 only if generator $i$ is connected to bus $n$ \cite{PKG-Matpower}. Node voltages $\bm{v}_n$ are expressed in polar coordinates as $|\bm{v}_n|\exp\left(j\bm{\theta}_n\right) \in \mathbb{C}^{|\mathcal{N}| \times 1}$, whereas complex power variables are in rectangular ones $\bm{s}=\bm{p}+j\bm{q}$.

The formulation builds upon PowerModels' \cite{PKG-PowerModels} standard one, with \eqref{eq-3B:kcl} and \eqref{eq-3B:power_flow} being \ac{KCL} and power flow equations. Inequalities \eqref{eq-3B:vm_limits}, \eqref{eq-3B:pg_limits}, \eqref{eq-3B:qg_limits}, \eqref{eq-3B:sb_limits} and \eqref{eq-3B:theta_limits} model the bounds for bus voltage magnitude, generator active and reactive power, branch apparent power and branch angle difference, respectively. It is modified simply by:
\begin{itemize}
    \item Replacing fixed complex load power in the \ac{KCL} equations \eqref{eq-3B:kcl} with the sum among the input load sample $\hat{\bm{s}}_d$ and the difference between the nonnegative load complex power slacks $\bm{s}_d^{\text{dw}}$ and $\bm{s}_d^{\text{up}}$.
    \item Adding box constraints \eqref{eq-3A:load_limits}, \eqref{eq-3A:qp_limits}, \eqref{eq-3A:pdtot_limit} to keep the complex load power \eqref{eq-3B:slack} inside the convex load polytope \eqref{eq-3A:polytope}. 
    \item Penalizing slack activation in the objective function \eqref{eq-3B:objective} using a linear cost coefficient $c_d$, such as the \ac{VOLL}, This must be significantly larger than the quadratic, linear and constant cost coefficients $\bm{c}_{g2}$, $\bm{c}_{g1}$ and $\bm{c}_{g0}$ for generator active power, to ensure that load slacks are used only if the sampled setpoint $\hat{\bm{s}}_d$ is unfeasible in the \ac{ACOPF} domain.    
\end{itemize}

By introducing load slack variables, the proposed formulation achieves a high convergence rate for feasible AC‑OPF instances, regardless of how infeasible the unclassified load space may be. However, extreme system loading scenarios, either too high or too low, can still yield no feasible solution due to constraint \eqref{eq-3A:pdtot_limit}. To address this, the first batch of samples in total active power load is used to identify and truncate those extremal regions in the uniform support \eqref{eq-3A:distribution} where convergence is consistently zero. Subsequent batches of samples are drawn from a nonparametric distribution by applying a truncated \ac{KDE} to previously converged system-wide active loading samples, with sampling probability inversely weighted by the \ac{KDE} density. In this way, sampling is focused on regions with limited, yet nonzero, convergence.

\begin{algorithm}[b]
    \caption{Procedure to generate an \ac{ACOPF} dataset}
    \label{alg:procedure}
    \begin{algorithmic}[1]
        \Require Grid model, $n_s$, $n_b$, $\delta_p$, $\delta_q$, $\delta_{\text{pf}}$, $\alpha_{\rm min}$, $\alpha_{\rm max}$, $\epsilon$
        \State Initialize polytope \eqref{eq-3A:polytope}: $\mathcal{P} \gets \{\bm{x} \mid \bm{A}\bm{x} \leq \bm{b}\}$
        \State Draw $\left[p_1^{\text{tot}}, \dots, p_{n_s}^{\text{tot}}\right]^\intercal \sim$ \eqref{eq-3A:distribution}
        \State $\mathcal{M}, \mathcal{C} \gets \emptyset$
        \For{$b = 1$ to $n_b$}
            \If{$b > 1$}
                \State Define distribution $\pi(p) \propto \frac{1}{\text{KDE}(\mathcal{C})(p) + \eta}$
                \State Draw $\left[p_1^{\text{tot}}, \dots, p_{n_s}^{\text{tot}}\right]^\intercal \sim \pi$
            \EndIf
            \For{$p_i^{\text{tot}}$ in $\left[p_1^{\text{tot}}, \dots, p_{n_s}^{\text{tot}}\right]$}
                \State Sample slice $\hat{\bm{s}}_d \sim \mathcal{U}(\mathcal{P}(p_i^{\text{tot}}))$
                \State Solve \eqref{eq-3B:opf}: $\overset{*}{m}_i \gets \text{OPF}(\hat{\bm{s}}_d)$
                \If{$f_{\text{is\_feasible}}(\overset{*}{m}_i)$}
                    \State $\mathcal{M} \gets \mathcal{M} \cup \{\overset{*}{m}_i\}$, $\mathcal{C} \gets \mathcal{C} \cup \{p_i^{\text{tot}}\}$
                \EndIf
            \EndFor
        \EndFor
        \State \textbf{return} dataset $\mathcal{M}$
    \end{algorithmic}
\end{algorithm}

The overall procedure to generate \ac{ACOPF} datasets is summarized in Alg. \ref{alg:procedure}, where $n_s$ is the number of target samples per batch and $n_b$ is the number of batches. Please note that, altogether, the only aspect modifying dataset quality with respect to OPFLearn \cite{GEN-Jones/OPFLearn} is the sampling in total active power load, a seemingly simple yet fundamental design choice. The algorithm is implemented in Julia language, using PowerModels \cite{PKG-PowerModels} and JuMP \cite{PKG-JuMP} libraries for power system optimization. Uniform sampling in the convex polytope is performed by relying on R's package Volesti \cite{VOL-Chalkis/Volesti}. While Alg. \ref{alg:procedure} exemplifies a serial implementation, in practice this is highly parallelized using Julia's distributed computing capabilities.

\section{Quality metrics}\label{sec:metrics}

As outlined in Section \ref{sec:intro}, the quality evaluation of an \ac{ACOPF} dataset can be performed using either quantitative metrics or empirical testing with \ac{ML} models. In this work, we focus exclusively on the former, as benchmarking datasets via \acp{NN} requires selecting a large number of hyperparameters and training configurations, which can significantly hinder results reproducibility and comparison. Accordingly, in this section we propose a qualitative, multi-criteria definition of quality of \ac{ACOPF} datasets for \ac{ML} applications, generated with \textit{generic} sampling methods. Then, existing evaluation measures are critically reviewed. As last, we introduce three novel, complementary metrics to assess quality quantitatively.

\subsection{Quality definition for AC-OPF datasets}

In the direct approximation paradigm (see \cite{NN-Zamzam/Learning, PINN-Fioretto/Dual, NN-Owerko/GNN, NN-Falconer/Leveraging, PINN-Donti/DC3, PINN-Nellikkath/Physics, NN-Singh/Sensitivity, PINN-Pan/DeepOPF, PINN-Liu/GNNDual, PINN-Owerko/GNN, PINN-Garcia/TypedGNN, PINN-Piloto/CANOS, PINN-Pareek/Bayesian, PINN-Park/PrimalDual, NN-Rahman2/Augmented, PINN-Yang/GuidedGNN}), a \ac{NN} model is trained to learn the mapping from input features to optimal \ac{ACOPF} regression variables whilst satisfying equality and inequality constraints of the problem. Within this general constrained optimization framework, we propose assessing the quality of an \ac{ACOPF} dataset for \ac{NN} training and testing based on the following three simple criteria: 

\begin{enumerate}
    \item High variance in primal \ac{ACOPF} regression variables.
    \item High complexity among \ac{ACOPF} instances.
    \item Frequent activation of variable bounds.
\end{enumerate}

The first two criteria both promote model generalization, preventing the \ac{NN} from learning trivial or overly simplistic patterns, and yet focus on different aspects of diversity. The first one concerns the variance of each individual primal variable across all instances in the dataset, ensuring broad range coverage and avoiding imbalanced distributions that can lead to large predictive errors in underrepresented regions \cite{SET-Yang/Imbalance}. In contrast, the second criterion examines complexity among AC-OPF instances, reflecting the degree of variation in operating conditions and diversity in interaction among variables. Finally, the third one targets constraint learning explicitly, since frequent activation of inequality limits across instances forces the \ac{NN} to learn constraint-aware representations. This is crucial to obtain trustworthy models for this safety-critical application.

To the best of our knowledge, no existing literature reference accounts for all these quality criteria. As outlined in Section \ref{sec:intro}, works \cite{GEN-Jones/OPFLearn, GEN-Ignasi/Rambo} and \cite{GEN-Ours} focus on the second one, measuring complexity through constraint activation. However, counting unique activation patterns \cite{GEN-Jones/OPFLearn} or tracking which constraints are activated at least once in the dataset \cite{GEN-Ignasi/Rambo}, \cite{GEN-Ours} does not quantify how much \ac{ACOPF} instances actually differ from one another. A more general notion of complexity is considered in \cite{GEN-Zhang/CSS} via the Simpson-diversity index, which increases if clusters in labeled datasets are numerous and balanced \cite{SET-Gregorius/Simpson}. Still, it ignores inter-cluster distances and, furthermore, its application to the \ac{ACOPF} requires clustering the instances artificially, making results less interpretable and dependent on hyperparameter choices. 

\subsection{Metric definition}

To address the identified research gap, this section proposes three novel metrics, each designed to quantify one of the quality criteria introduced above. Let $K$ denote the target number of \ac{ACOPF} instances in the dataset, $t$ a given variable type (e.g., generator active power) and $\mathcal{T}$ the set to which it belongs. Furthermore, let $\bm{M}_t \in \mathbb{R}^{K \times |\mathcal{T}|}$ be a generic matrix containing in each row the variable results for an instance. Then, the binary matrix $\bm{A}_{t_2} \in \{0,1\}^{K \times l|\mathcal{T}|}$ encodes the activation status of the $l$ variable limits, where different limits are treated separately by horizontally concatenating their activation indicators. Differently, in $\bm{A}_{t_3} \in \{-1,0,1\}^{K \times |\mathcal{T}|}$ lower and upper bounds, or, equivalently, bounds at the \textit{from} and \textit{to} ends of a branch, are considered as mutually exclusive and are thus stored within a single matrix entry.

Within this framework, variance in primal regression variable $t$ (i.e., first criterion) is measured through the mean normalized Shannon entropy:

\begin{equation}\label{eq-4:q1}
    Q_{1,t} = \frac{1}{|\mathcal{T}|}\sum_{i=1}^{|\mathcal{T}|} \frac{-\sum_{b=1}^B \pi_{i,b}\log_2 \pi_{i,b}}{\log_2 B} \in \left[0,1\right]
\end{equation}
where $B$ is the number of bins (e.g., 100) used to discretize each column of $\bm{M}_t$ and $\pi_{i,b}$ is the empirical probability of the \( b \)-th bin for column $i$. This assesses, on average, how uniformly data is distributed across its feasible range.

For the second criterion, we build on the concept of constraint activation from \cite{GEN-Jones/OPFLearn, GEN-Ignasi/Rambo}, extending it to directly quantify diversity among activation patterns. Specifically, instance complexity is measured by the average normalized Hamming distance across all pairs of ternary activation patterns:

\begin{equation}\label{eq-4:q2}
    Q_{2,t} = \frac{1}{\binom{K}{2}}\sum_{1 \leq i < j \leq K } \frac{\|\bm{1} ( \bm{A}_{i,:} \neq \bm{A}_{j,:}) \|_1}{|\mathcal{T}|} \in \left[0,1\right]
\end{equation}
where $\bm{A}$ is the three-state matrix $\bm{A}_{t_3}$. Practically, this metric counts, on average, by how many states two instances differ for variable $t$, with low values arising either because the variable is fixed at given bound or because bounds are never active.

The last measure, which focuses on the third criterion, helps differentiating between these two cases. This simply computes the average normalized bound activation frequency over the subset of limits $\mathcal{L}_{nr} = \left\{ \bm{a}_i \in \bm{A}_{t_2} \mid \bm{a}_i^\intercal \bm{1} > 0 \right\}$ that are activated at least once in the dataset:

\begin{equation}\label{eq-4:q3}
    Q_{3,t} = \frac{1}{|\mathcal{L}_{nr}^{\rm \max}|}\sum_{a_i \in \mathcal{L}_{nr}} \min \left( \frac{\bm{a}_i^\intercal \bm{1}}{K}, \frac{1}{2} \right) \in \left[0,0.5\right]
\end{equation}
where the normalized frequency of a given bound is clipped at 0.5 to favor balanced activations between lower and upper limits. Overall, this allows focusing on non-redundant constraints \cite{OPF-Aquino/Redundant}, verifying if they are well-represented and thus learnable by the \ac{NN}. Since set cardinality $|\mathcal{L}_{nr}|$ may vary across different dataset generation methods \cite{GEN-Ignasi/Rambo}, the metric is averaged by the largest set of non-redundant constraints $|\mathcal{L}_{nr}^{\rm max}|$ among the tested approaches, penalizing those with limited complexity.

Overall, the three metrics are designed to be highly informative, easily interpretable, scalable and complementary, in the sense that each captures a distinct quality aspect, while their combined interpretation yields additional insights.

\section{Methodology comparison}\label{sec:comparison}

\begin{table*}[!b]
    \begin{minipage}{0.62\textwidth}
        \vspace{-1.0em}
        \centering
        \refstepcounter{table}
        \caption{Metric $Q_1$ results in percentage for multiple methods and test cases, with sub-tables (a)-(d) corresponding to variables $\bm{p}_g$, $\bm{q}_g$, $|\bm{v}_n|$, $\bm{\theta}_n$, respectively. Bold indicates the best performing method for each grid and variable.}
        \label{tab:q1}
        \begin{tabular}{@{}cc@{}}
            \begin{tabular}{c ccccc}
                & \multicolumn{5}{c}{(a)} \vspace{0.5em} \\
                \toprule
                $|\mathcal{N}|$ & M0/20 & M0   & M1   & M2   & MX   \\
                \midrule
                30    & 35.1  & 42.5 & 51.7 & \textbf{51.7} & 49.7          \\
                39    & 13.9  & 20.2 & 22.6 & 25.6          & \textbf{27.7} \\
                57    & 21.7  & 33.4 & 13.7 & \textbf{33.8} & 32.5          \\
                118   & 10.8  & 18.1 & 19.3 & 21.7          & \textbf{24.7} \\
                500   & 3.2   & 6.8  & N/A  & N/A           & \textbf{15.6} \\
                1354  & 2.3   & 9.3  & N/A  & N/A           & \textbf{14.9} \\
                2000  & 6.2   & 12.1 & N/A  & N/A           & \textbf{20.2} \\
                2869  & 2.8   & 6.1  & N/A  & N/A           & \textbf{12.4} \\
                4661  & 5.2   & 13.7 & N/A  & N/A           & \textbf{19.7} \\
                \bottomrule
            \end{tabular} &
            \begin{tabular}{ccccc}
                \multicolumn{5}{c}{(b)} \vspace{0.5em} \\
                \toprule
                M0/20 & M0   & M1   & M2   & MX   \\
                \midrule
                31.4 & 45.7          & \textbf{48.7} & 39.7          & 46.9 \\
                29.2 & 43.7          & \textbf{50.0} & 42.3          & 41.8 \\
                23.5 & 35.1          & \textbf{43.6} & 39.5          & 40.5 \\
                20.3 & 35.2          & 31.9          & \textbf{40.3} & 40.3 \\
                26.2 & \textbf{40.1} & N/A           & N/A           & 37.2 \\
                22.7 & \textbf{37.1} & N/A           & N/A           & 34.1 \\
                23.2 & 37.6          & N/A           & N/A           & \textbf{37.9} \\
                22.7 & 35.3          & N/A           & N/A           & \textbf{35.9} \\
                27.1 & \textbf{43.2} & N/A           & N/A           & 43.0 \\
                \bottomrule
            \end{tabular}      
        \end{tabular}
        
        \vspace{1em}
        
        \begin{tabular}{@{}cc@{}}
            \begin{tabular}{c ccccc}
                & \multicolumn{5}{c}{(c)} \vspace{0.5em} \\
                \toprule
                $|\mathcal{N}|$ & M0/20 & M0   & M1   & M2   & MX   \\
                \midrule
                30    & 27.8  & 51.0 & 50.1 & \textbf{56.8} & 53.4          \\
                39    & 40.8  & 55.9 & 52.8 & \textbf{59.5} & 59.1          \\
                57    & 26.5  & 49.0 & 50.4 & \textbf{58.7} & 56.2          \\
                118   & 28.2  & 49.2 & 51.3 & 57.3          & \textbf{58.2} \\
                500   & 18.8  & 45.4 & N/A  & N/A           & \textbf{59.6} \\
                1354  & 24.7  & 55.1 & N/A  & N/A           & \textbf{57.2} \\
                2000  & 23.0  & 45.3 & N/A  & N/A           & \textbf{60.4} \\
                2869  & 24.3  & 44.4 & N/A  & N/A           & \textbf{56.5} \\
                4661  & 26.5  & 55.0 & N/A  & N/A           & \textbf{58.0} \\
                \bottomrule
            \end{tabular} &
            \begin{tabular}{ccccc}
                \multicolumn{5}{c}{(d)} \vspace{0.5em} \\
                \toprule
                M0/20 & M0   & M1   & M2   & MX   \\
                \midrule
                11.6  & 31.4 & 27.3 & \textbf{45.1} & 43.5          \\
                23.3  & 40.3 & 35.7 & \textbf{47.0} & 46.5          \\
                12.0  & 35.2 & 21.2 & \textbf{42.8} & 38.3          \\
                25.6  & 45.2 & 40.3 & \textbf{54.4} & 51.4          \\
                19.2  & 39.1 & N/A  & N/A           & \textbf{46.1} \\
                22.2  & 47.5 & N/A  & N/A           & \textbf{52.6} \\
                11.7  & 34.8 & N/A  & N/A           & \textbf{47.3} \\
                29.3  & 47.4 & N/A  & N/A           & \textbf{57.2} \\
                16.5  & 34.7 & N/A  & N/A           & \textbf{43.0} \\
                \bottomrule
            \end{tabular}       
        \end{tabular}
    \end{minipage}
    \hspace{0.25em}
    \begin{minipage}{0.34\textwidth}
        \centering
        \setlength{\abovecaptionskip}{1pt}
        \includegraphics[width=0.91\linewidth]{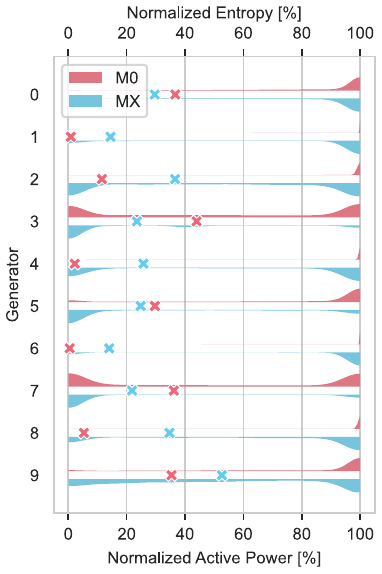}
        \centering
        \captionof{figure}{\ac{KDE} distribution of min-max scaled active power values for generators of the 39-bus system (lower x-axis) against corresponding Shannon entropy values (upper x-axis) for selected methods.}
        \label{fig:q1-kde}
    \end{minipage}
\end{table*}

This section evaluates the quality of AC-OPF datasets generated using the methodology from Section \ref{sec:method}, based on the metrics defined in \ref{sec:metrics}. A robust comparison is carried out against three existing approaches, namely independent uniform random sampling, OPFLearn \cite{GEN-Jones/OPFLearn}, and RAMBO \cite{GEN-Ignasi/Rambo}, alongside datasets from the OPFData repository \cite{GEN-Lovett/OPFData}, which serve as a reference for uniform sampling with a widely-used variation range (i.e., $\pm 20\%$). The comparison structure is first detailed, followed by the presentation of its results, with a final brief focus of computational performance and scalability.

\subsection{Comparison structure}\label{subsec:structure}

The comparison among \ac{ACOPF} dataset generation methods is designed to be robust and reliable based on three key principles. First, to guarantee fairness, all approaches, except OPFData \cite{GEN-Lovett/OPFData}, are tested under consistent assumptions regarding boundaries of the unclassified input load space. As shown in Table \ref{tab:load_space}, most methods differ exclusively by the $q_d/p_d$ ratio, which reduces the sampling space, with only M1 operating on a wider reactive variation range. Second, extensive testing is performed on nine power system test cases from the PGLib library v21.07 \cite{OPF-Babaeine/PGLib}, namely the IEEE 30, 57 and 118, EPRI 39, GOC 500 and 2000, PEGASE 1354 and 2869 and SDET 4661-bus systems. Due to scalability issues, however, OPFLearn and RAMBO are applied only up to the IEEE 118-bus case. Third, to enable a detail assessment, metrics are evaluated seperately per variable. In particular, $Q_1$ \eqref{eq-4:q1} is computed for all \ac{ACOPF} primal regression variables, namely generator active/reactive power and bus voltage magnitude/phase angle, while metrics $Q_2$ \eqref{eq-4:q2} and $Q_3$ \eqref{eq-4:q3} are applied to inequality constraints \eqref{eq-3B:vm_limits}, \eqref{eq-3B:pg_limits}, \eqref{eq-3B:qg_limits} and \eqref{eq-3B:sb_limits}.

\begin{table}[ht]
    \centering
    \vspace{-0.5em}
    \addtocounter{table}{-2}
    \caption{Assumptions on unclassified input load space}
    \label{tab:load_space}
    \begin{tabular}{llccc}
        \toprule
         Method & Reference & $\delta_p$ & $\delta_q$ & $q_d/p_d$ ratio  \\
         \midrule
         M0/20  & OPFData \cite{GEN-Lovett/OPFData}  & 20  & 20     & N \\
         M0     & Uniform sampling                   & 100 & 100    & N \\
         M1     & OPFLearn \cite{GEN-Jones/OPFLearn} & 100 & $>100$ & Y \\
         M2     & RAMBO \cite{GEN-Ignasi/Rambo}      & 100 & 100    & N \\
         MX     & HEDGeOPF (our)                     & 100 & 100    & Y \\
         \bottomrule
    \end{tabular}
    \refstepcounter{table}
\end{table}

Each method listed in Table \ref{tab:load_space} is used to generate \ac{ACOPF} datasets of 10000 instances for the selected test cases, which are identified by bus number $|\mathcal{N}|$. For the OPFData repository (i.e., M0/20), which provides datasets of 300000 instances, the first 10000 are used. The missing datasets for M0/20, as well as those for M0, are generated using formulation \eqref{eq-3B:opf} with constraints \eqref{eq-3A:qp_limits} and \eqref{eq-3A:pdtot_limit} removed. In contrast, datasets for M1 and M2 are obtained using their original open-source implementations \cite{PKG-OPFLearn} and \cite{PKG-RAMBO}. Lastly, HEDGeOPF's power factor parameters $\delta_{\text{pf}}$, $\alpha_{\rm max}$ and $\alpha_{\rm min}$ are fixed to 0.05, 0.99 and 0.01, respectively, throughout all simulations.

\subsection{Simulation results}\label{subsec:results}

Table \ref{tab:q1} reports the results in percentage for metric $Q_1$ \eqref{eq-4:q1}, with Fig. \ref{fig:q1-kde} exemplifying what different values of normalized Shannon entropy actually mean in terms of data distribution.
In relative terms, both M2 and MX methods exhibit a consistent, stable increase in $Q_1$ for generator active power, voltage magnitude and phase angle, with MX improving over M0 on average by 55\%, 16\% and 21\%, respectively. However, when comparing M2 and MX, the former generally achieves slightly better results, particularly for voltage variables, with average relative gain of 2.4\% in magnitude and 5.6\% in phase angle. The only exception is generator reactive power, for which no clear winner emerges. Notably, M1 performs best on the smaller test cases of Table \ref{tab:q1}-(b) most likely due to the broader range of variation reported in Table \ref{tab:load_space}. For the remaining variables, instead, M1 scores on par with M0.

\setlength{\abovecaptionskip}{1pt}
\begin{figure*}[!b]
    \vspace{-0.6em}
    \centering
    \includegraphics[width=0.825\textwidth]{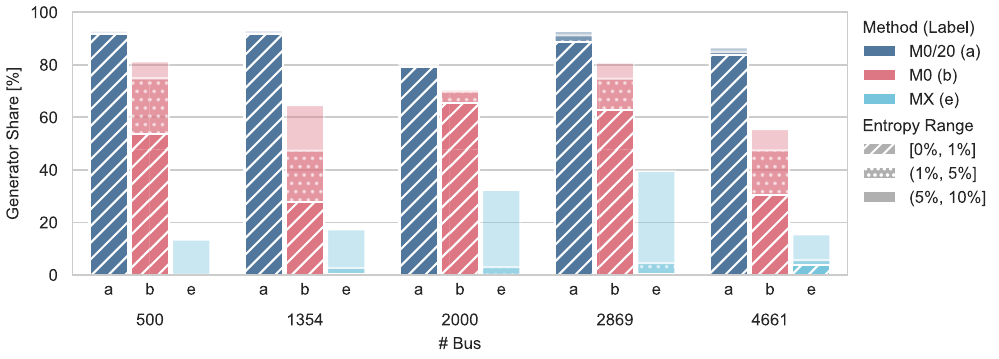}
    \caption{Share of generators with normalized Shannon entropy in active power values below 10\%, with partition into three entropy sub-intervals. Results are reported for the five largest test cases and for selected methods, which are labeled according to the order defined in Table \ref{tab:load_space}.}
    \label{fig:q1-share}
\end{figure*}

In absolute terms, the mean normalized entropy $Q_1$ of generator reactive power, bus voltage magnitude and phase angle is sufficiently high and stable across system sizes for all methods except M0/20. This implies that values for these variables are well spread over their feasible ranges, reducing the risk of \acp{NN} memorizing overly simplistic trends. In contrast, generator active power $Q_1$ values in Table \ref{tab:q1}-(a) are significantly lower and tend to decrease with system size, particularly for M0-class methods.
This behavior aligns with the theoretical expectations discussed in Section \ref{sec:problem}: minimizing generation cost, as in \eqref{eq-3B:opf}, causes generators to operate predominantly at their lower or upper limits, except when marginal or constrained by network topology. Consequently, as shown in Fig. \ref{fig:q1-kde}, the intermediate region of the active power distribution tends to be inherently underrepresented for most generators, regardless of the sampling method. Distribution skewness, however, is further exacerbated in approaches that lack explicit control over total system loading (i.e., M0/20, M0 and M1), thereby limiting its variability and restricting the set of generators that can be marginal to only a few. In such cases, active power distributions can become nearly degenerate for those generators, such as units 1, 2, 4, 6 and 8 for M0 in Fig. \ref{fig:q1-kde}, that always work at one of their bounds. As illustrated in Fig. \ref{fig:q1-kde} for MX, this issue can be effectively mitigated by increasing variability in total active power loading, either through structured random sampling (MX) or optimization-based exploration (M2).

Since this limitation worsens with system size for the M0-class methods of Table \ref{tab:q1}-(a), Fig. \ref{fig:q1-share} reports the share of generators exhibiting a nearly degenerate univariate distribution in the five largest test cases. Three levels of degeneracy are distinguished based on normalized entropy thresholds. Overall, at least 80\% and 50\% of generators in M0/20 and M0 datasets, respectively, display highly degenerate distributions (i.e., entropy below 5\%) in active power. This indicates that \acp{NN} trained on such datasets are prone to learn trivial mappings between load inputs to fixed generator setpoints, severely limiting their generalization capabilities. As shown in Fig. \ref{fig:q1-share}, MX effectively mitigates this issue.

\begin{figure*}[!t]
    \centering
    \includegraphics[width=0.825\textwidth]{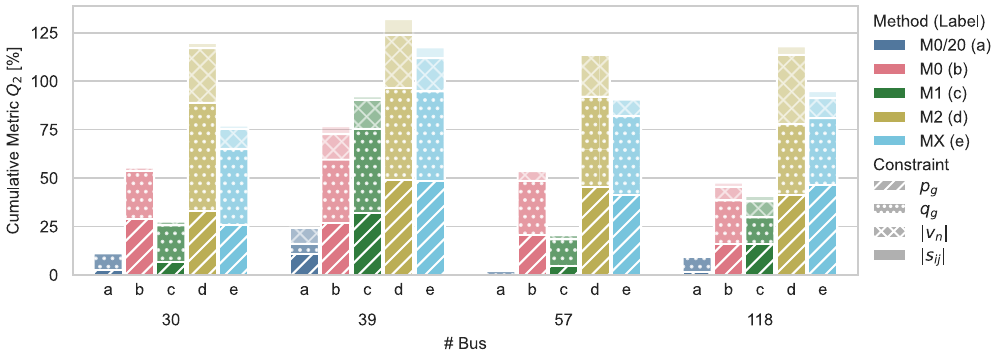}
    \caption{Comparison among every method of Table \ref{tab:load_space} in terms cumulative mean normalized Hamming distance (i.e., $Q_2$ metric \eqref{eq-4:q2}) over the different constraint classes. Results are reported for the four smallest test cases.}
    \label{fig:q2}
\end{figure*}

The results for metrics $Q_2$ further highlight the superiority of M2 and, as second best, MX over the other methods. In particular, Fig. \ref{fig:q2} displays the cumulative $Q_2$ percentage values over the considered inequality constraints up to the 118-bus system, whereas results for larger test cases are reported in Table \ref{tab:q2}. Following the approach of \cite{GEN-Ignasi/Rambo}, a bound is labeled as active if the corresponding variable lies within 1\% of its feasible range from that bound.

On average, MX yields substantial improvements over M0, specifically 254\%, 60\%, 87\%, and 107\% in generator active and reactive power, bus voltage magnitude, and branch apparent power constraints, respectively. However, while MX is competitive with M2 on metric $Q_1$, the results for $Q_2$ make M2's superiority much more evident. As shown in Fig. \ref{fig:q2}, the most pronounced difference arises in bus voltage magnitude, where M2 achieves a 157\% average improvement over MX. This reflects the core design of M2, which explicitly targets the edges of the \ac{ACOPF} feasible space by maximizing distance in key variables, such as optimal bus voltage magnitude, across different instances \cite{GEN-Ignasi/Rambo}. A similar trend holds for branch apparent power constraints, where M2 achieves 45\% higher complexity than MX. In contrast, the two methods perform nearly identically on generator active power (see Fig. \ref{fig:q2}), strongly suggesting that MX’s simple sampling heuristic is highly effective at enhancing $Q_2$ complexity in this dimension.

For the remaining methods, it is possible to detect additional relevant behaviors:
\begin{itemize}
    \item Fig. \ref{fig:q2} confirms that M1 is not superior to M0. In particular, the high entropy in generator reactive power of Table \ref{tab:q1}-(b) does not translate into greater diversity in constraint activation patterns.
    \item M0's low complexity in generator active power, as shown in Table \ref{tab:q2}-(a), strongly correlates with degeneracy of marginal distributions in Fig. \ref{fig:q1-share}.
    \item M0/20 datasets exhibit minimal $Q_2$ complexity across all tested cases, especially for branch apparent power. This further suggests that \acp{NN} trained on these datasets may simply memorize the limited set of interaction patterns.
\end{itemize}
    
As last remark, it can be clearly seen from Fig. \ref{fig:q2} and Table \ref{tab:q2} that, except for M0/20, $Q_2$ complexity is in general significantly larger for generator active and reactive power than for bus voltage magnitude and branch apparent power. Notably, $Q_2$ values for the latter are close to zero in the IEEE 57 and GOC 500 and 2000-bus systems.

\setlength{\abovecaptionskip}{1pt}

\begin{table}[!t]
    \vspace{-1.0em}
    \centering
    \caption{Metric $Q_2$ results in percentage for multiple methods and test cases, with sub-tables (a)-(d) corresponding to constraints \eqref{eq-3B:pg_limits}, \eqref{eq-3B:qg_limits}, \eqref{eq-3B:vm_limits}, \eqref{eq-3B:sb_limits}, respectively.}
    \label{tab:q2}
    \begin{tabular}{@{}cc@{}}
        \begin{tabular}{c ccc}
            & \multicolumn{3}{c}{(a)} \vspace{0.5em} \\
            \toprule
            $|\mathcal{N}|$ & M0/20 & M0 & MX   \\
            \midrule
            500   & 1.0   & 9.4  & \textbf{47.4} \\
            1354  & 2.6   & 19.2 & \textbf{44.8} \\
            2000  & 0.4   & 5.6  & \textbf{44.5} \\
            2869  & 2.1   & 9.6  & \textbf{39.9} \\
            4661  & 2.8   & 20.2 & \textbf{43.4} \\
            \bottomrule
        \end{tabular} &
        \begin{tabular}{ccc}
            \multicolumn{3}{c}{(b)} \vspace{0.5em} \\
            \toprule
            M0/20 & M0 & MX   \\
            \midrule
            7.4   & 33.2 & \textbf{51.0} \\
            6.4   & 39.3 & \textbf{47.6} \\
            2.5   & 15.0 & \textbf{39.5} \\
            8.6   & 23.8 & \textbf{43.6} \\
            6.8   & 26.6 & \textbf{29.8} \\
            \bottomrule
        \end{tabular}
    \end{tabular}
    
    \vspace{1em}
    
    \begin{tabular}{@{}cc@{}}
        \begin{tabular}{c ccc}
            & \multicolumn{3}{c}{(c)} \vspace{0.5em} \\
            \toprule
            $|\mathcal{N}|$ & M0/20 & M0 & MX   \\
            \midrule
            500   & 1.4   & \textbf{5.1}  & 4.7           \\
            1354  & 2.5   & 9.9           & \textbf{11.8} \\
            2000  & 0.4   & 1.2           & \textbf{2.1}  \\
            2869  & 2.1   & 5.8           & \textbf{8.0}  \\
            4661  & 1.3   & 6.3           & \textbf{6.7}  \\
            \bottomrule
        \end{tabular} &
        \begin{tabular}{ccc}
            \multicolumn{3}{c}{(d)} \vspace{0.5em} \\
            \toprule
            M0/20 & M0 & MX   \\
            \midrule
            0.01  & 0.29 & \textbf{0.86} \\
            0.09  & 1.43 & \textbf{2.87} \\
            0.00  & 0.05 & \textbf{0.13} \\
            0.14  & 0.64 & \textbf{1.27} \\
            0.51  & 4.56 & \textbf{7.19} \\
            \bottomrule
        \end{tabular}
    \end{tabular}
\end{table}

\begin{table}[!t]
    \vspace{-1.0em}
    \centering
    \caption{Metric $Q_3$ results [\%], with sub-tables organized as in Table \ref{tab:q2}. Two-level cell shading indicates for each method in each sub-table the relative size of non-redundant constraint set when below 40\% (darker is [0, 20]\%). Underlined methods have the largest set.}
    \label{tab:q3}
    \begin{tabular}{@{}cc@{}}
        \begin{tabular}{c cccc}
            & \multicolumn{4}{c}{(a)} \vspace{0.5em} \\
            \toprule
            $|\mathcal{N}|$ & M0   & M1   & M2   & MX   \\
            \midrule
            30    & \lc{18.7} & \lc{2.4}  & \underline{\textbf{20.3}} & \underline{18.3}          \\
            39    & 22.1      & 24.3      & \underline{\textbf{34.2}} & \underline{31.3}          \\
            57    & 14.6      & 16.0      & \underline{\textbf{28.6}} & 25.7                      \\
            118   & 20.4      & 18.1      & \underline{31.9}          & \textbf{32.4}             \\
            500   & 22.6      & N/A       & N/A                       & \underline{\textbf{41.1}} \\
            1354  & 27.2      & N/A       & N/A                       & \underline{\textbf{39.5}} \\
            2000  & \lc{12.3} & N/A       & N/A                       & \underline{\textbf{34.2}} \\
            2869  & 20.1      & N/A       & N/A                       & \underline{\textbf{37.6}} \\
            4661  & 24.9      & N/A       & N/A                       & \underline{\textbf{35.5}} \\
            \bottomrule
        \end{tabular} &
        \begin{tabular}{cccc}
            \multicolumn{4}{c}{(b)} \vspace{0.5em} \\
            \toprule
            M0   & M1   & M2   & MX   \\
            \midrule
            10.3 & 8.8              & \underline{\textbf{28.7}} & 17.7                      \\
            14.2 & \underline{18.9} & \underline{\textbf{23.4}} & 21.5                      \\
            19.6 & 7.8              & \underline{\textbf{27.0}} & \underline{25.3}          \\
            16.7 & 15.3             & \underline{\textbf{25.3}} & 22.6                      \\
            19.8 & N/A              & N/A                       & \underline{\textbf{29.1}} \\
            22.1 & N/A              & N/A                       & \underline{\textbf{28.3}} \\
            9.6  & N/A              & N/A                       & \underline{\textbf{22.1}} \\
            15.3 & N/A              & N/A                       & \underline{\textbf{25.4}} \\
            18.4 & N/A              & N/A                       & \underline{\textbf{22.5}} \\
            \bottomrule
        \end{tabular}      
    \end{tabular}
    
    \vspace{1em}
    
    \begin{tabular}{@{}cc@{}}
        \begin{tabular}{c cccc}
            & \multicolumn{4}{c}{(c)} \vspace{0.5em} \\
            \toprule
            $|\mathcal{N}|$ & M0   & M1   & M2   & MX   \\
            \midrule
            30    & \mc{1.9}          & \dc{1.9}  & \underline{\textbf{10.1}}      & \mc{2.7}                      \\
            39    & \lc{4.9}          & \lc{5.1}  & \underline{\textbf{9.8}}       & 4.7                           \\
            57    & \mc{2.2}          & \mc{0.9}  & \lc{\underline{\textbf{10.9}}} & \mc{4.1}                      \\
            118   & \lc{2.3}          & \lc{2.8}  & \underline{\textbf{12.6}}      & 2.7                           \\
            500   & \mc{\textbf{3.4}} & N/A  & N/A                                 & \mc{\underline{2.7}}          \\
            1354  & \lc{4.0}          & N/A  & N/A                                 & \lc{\underline{\textbf{5.0}}} \\
            2000  & \dc{1.4}          & N/A  & N/A                                 & \dc{\underline{\textbf{2.2}}} \\
            2869  & \mc{3.3}          & N/A  & N/A                                 & \lc{\underline{\textbf{4.1}}} \\
            4661  & \mc{4.7}          & N/A  & N/A                                 & \mc{\underline{\textbf{5.5}}} \\
            \bottomrule
        \end{tabular} &
        \begin{tabular}{cccc}
            \multicolumn{4}{c}{(d)} \vspace{0.5em} \\
            \toprule
            M0   & M1   & M2   & MX   \\
            \midrule
            \dc{\underline{8.8}} & \dc{\underline{10.0}} & \dc{\underline{\textbf{14.8}}} & \dc{\underline{8.3}}          \\
            \mc{2.3}             & \dc{1.1}              & \lc{\underline{\textbf{5.4}}}  & \mc{3.4}                      \\
            \dc{0.1}             & \dc{0.0}              & \dc{\underline{\textbf{1.3}}}  & \dc{0.0}                      \\
            \dc{2.4}             & \dc{3.9}              & \mc{\underline{\textbf{5.4}}}  & \dc{3.4}                      \\
            \dc{0.9}             & N/A              & N/A                       & \dc{\underline{\textbf{1.8}}}           \\
            \dc{4.1}             & N/A              & N/A                       & \dc{\underline{\textbf{6.4}}}           \\
            \dc{0.3}             & N/A              & N/A                       & \dc{\underline{\textbf{0.9}}}           \\
            \dc{4.7}             & N/A              & N/A                       & \dc{\underline{\textbf{7.0}}}           \\
            \mc{6.7}             & N/A              & N/A                       & \mc{\underline{\textbf{9.8}}}           \\
            \bottomrule
        \end{tabular}       
    \end{tabular}
\end{table}

The results for metric $Q_3$, which are reported as percentage in Table \ref{tab:q3}, help explain this behavior. As shown in Table \ref{tab:q3}-(c) and Table \ref{tab:q3}-(d), the limited $Q_2$ complexity in bus voltage magnitude and branch apparent power stems from a low proportion of non-redundant constraints, indicated by cell shading, and from the fact that these are rarely binding across the instances. As a result, regardless of the dataset generation method, \acp{NN} trained on the \ac{ACOPF} problem  are likely to neglect operational patterns where these bounds become critical, as they are underrepresented in the training data. This is especially true if the penalty loss for constraint violations is weighted by the size of the full constraint set rather than the reduced, non-redundant one, an approach still widespread in the literature. In addition, datasets with low activation frequency should be used with caution when assessing feasibility of \ac{NN} predictions at testing, to avoid overly optimistic conclusions. For example, achieving a high average feasibility rate in branch apparent power in the IEEE 57-bus system, which is a highly used test case, cannot be regarded as empirical proof that the \ac{NN} has truly learned to abide by this constraint class.

The remaining bounds of Table \ref{tab:q3}-(a) and Table \ref{tab:q3}-(b) display sufficiently high activation frequencies, which should enable \acp{NN} to learn constraint-aware representations, provided that corresponding marginal distributions are not degenerate. Altogether, $Q_3$ confirms the relative performance differences among methods already observed through metrics $Q_1$ and $Q_2$.

Overall, the dataset comparison provides empirical validation of the theoretical expectations outlined in Section \ref{sec:problem}. Specifically, it confirms that M1 does not improve at all over M0 and that the quality of M0-class datasets is significantly degraded, particularly in generator active power distributions, with M0/20 showing almost no $Q_2$ complexity in any variable class. Furthermore, it highlights that MX's uniform sampling in total active power load is effective at increasing diversity in \ac{ACOPF} solutions, not only in terms of marginal distributions, but also with respect to complexity in variable interactions and constraint activation frequency. Still, the comparison also shows that MX is consistently outperformed by M2, especially under metrics $Q_2$ and $Q_3$.

\subsection{Computational performance}\label{subsec:timing}

The technical superiority of M2 (i.e. RAMBO) comes, however, at a significant computational cost that currently hinders its scalability to large-scale power systems. To this end, Table \ref{tab:solve_time} compares, for multiple test cases, the average solve times across \ac{ACOPF} feasible instances for the most computationally intensive step in M2 and MX. In the former, this consist in a warm-started, bilevel optimization problem with non-convex objective function and relaxed complementary slackness constraints \cite{GEN-Ignasi/Rambo}, whereas in the latter it is the \ac{ACOPF} formulation \eqref{eq-3B:opf}. As shown in Table \ref{tab:solve_time}, M2's solve time is between 30 and 115 times larger than MX's one, scaling with system size. Unfortunately, it was not possible to obtain this information for M2 on the larger test cases due to runtime errors. Nevertheless, it is still worth noting that MX's solve time only exceeds that of M2 on the 118-bus system when scaling up to the 4661-bus grid, which requires 144 seconds on average per \ac{ACOPF} instance.

\begin{table}[ht]
    \centering
    \caption{Ipopt average solve time [s] for the most intensive computational step in M2 and MX for multiple grids.}
    \label{tab:solve_time}
    \begin{tabular}{lcccc}
        \toprule
        \multirow{2}{*}{Method} & \multicolumn{4}{c}{$|\mathcal{N}|$} \\
        \cmidrule(rl){2-5}
        & 30 & 39 & 57 & 118 \\
        \midrule
        M2 & 3.40 & 18.45 & 12.25 & 92.49 \\
        MX & 0.11 & 0.28 & 0.23 & 0.81 \\
        \bottomrule
    \end{tabular}
\end{table}
All simulations are run on a machine equipped with two Intel\textsuperscript{\textregistered}~Xeon\textsuperscript{\textregistered}~Gold 6248R processors and a total of 128 GB of RAM. IPOPT \cite{IPOPT} is used as \ac{NLP} solver with MUMPS as the default linear solver.

\section{Conclusion}\label{sec:conclusions}

This work tackles two central challenges in generating high-quality \ac{ACOPF} datasets for \ac{ML} applications: (i) enhancing method scalability and dataset quality beyond the current state-of-the-art, and (ii) establishing a principled, quantitative framework for dataset quality assessment that enables fair and reproducible benchmarking across methods.

To address the first objective, we introduce a scalable dataset generation heuristic that directly samples the input, unclassified convex load space by solving an \ac{ACOPF} formulation augmented with load slack variables. This approach fully eliminates the need for preemptive volume reduction strategies aimed at improving convergence, such as those adopted in \cite{GEN-Venzke/Efficient, GEN-Jones/OPFLearn} and \cite{GEN-Ours}. In addition, sampling is not performed uniformly in the load convex polytope, in order to avoid the concentration of measure effect common in high dimensional spaces. Instead, it is carried out uniformly with respect to total load active power, thereby increasing variability along this key dimension.

To assess \ac{ACOPF} dataset quality for \ac{ML} applications, we define three complementary quality criteria and propose robust, informative metrics to quantify each. The first two focus on dataset complexity, with $Q_1$ capturing variability in the marginal distributions of AC-OPF primal variables, and $Q_2$ measuring the diversity in constraint activation patterns among instances, respectively. The third metric, $Q_3$, targets constraint-aware learning by quantifying the activation frequency of variable bounds.

These metrics are then used to compare our approach, MX, against three open-source dataset generation methods, namely independent uniform random sampling (M0), OPFLearn \cite{GEN-Jones/OPFLearn} (M1) and RAMBO \cite{GEN-Ignasi/Rambo} (M2), as well as the OPFData repository \cite{GEN-Lovett/OPFData} (M0/20), yielding several key insights. Notably, uniform sampling in total load active power significantly increases $Q_1$ and $Q_2$ complexities across all variable types. This explains why MX consistently outperforms M0 and M1 on all test cases, also confirming that M1’s limited effectiveness stems primarily from the concentration of measure effect. In contrast, MX remains consistently outmatched by M2, especially in terms of $Q_2$ and $Q_3$ metrics on bus voltage magnitude. It must be noted, however, that all methods display a low activation rate for bus voltage magnitude and branch apparent power constraints, thereby suggesting that \ac{ML} models may struggle at learning such bounds, regardless of the dataset generation approach.

Finally, the comparison provides a detailed characterization for low quality in \ac{ACOPF} datasets generated with M0/20, the most utilized method in the literature. Specifically, M0/20 produces datasets with a high share of generators exhibiting degenerate active power distributions and almost no diversity in bound activation patterns across any variable class. This raises serious concerns about the reliability of \ac{ML} models trained and tested on such datasets, stressing the need to move beyond the M0/20 method for \ac{ML}-oriented research on the \ac{ACOPF} problem.

Overall, our heuristic HEDGeOPF emerges as a practical and robust choice, especially for large-scale power systems or extensive datasets, as it achieves the best balance between quality and scalability among all tested methods. Nonetheless, the persistent performance gap relative to RAMBO indicates that there is still room for improving \ac{ACOPF} dataset quality.

\bibliographystyle{IEEEtran}

\begin{thebibliography}{10}
\providecommand{\url}[1]{#1}
\csname url@samestyle\endcsname
\providecommand{\newblock}{\relax}
\providecommand{\bibinfo}[2]{#2}
\providecommand{\BIBentrySTDinterwordspacing}{\spaceskip=0pt\relax}
\providecommand{\BIBentryALTinterwordstretchfactor}{4}
\providecommand{\BIBentryALTinterwordspacing}{\spaceskip=\fontdimen2\font plus
\BIBentryALTinterwordstretchfactor\fontdimen3\font minus \fontdimen4\font\relax}
\providecommand{\BIBforeignlanguage}[2]{{%
\expandafter\ifx\csname l@#1\endcsname\relax
\typeout{** WARNING: IEEEtran.bst: No hyphenation pattern has been}%
\typeout{** loaded for the language `#1'. Using the pattern for}%
\typeout{** the default language instead.}%
\else
\language=\csname l@#1\endcsname
\fi
#2}}
\providecommand{\BIBdecl}{\relax}
\BIBdecl

\bibitem{GEN-Ours}
L.~Perbellini, M.~Baù, and S.~Grillo, ``{An Efficient and Scalable Algorithm for the Creation of Representative Synthetic AC-OPF Datasets},'' in \emph{2024 IEEE 8th Forum on Research and Technologies for Society and Industry Innovation (RTSI)}, 2024, pp. 653--658.

\bibitem{OPF-Carpentier(1962)}
J.~Carpentier, ``Contribution \'{a} l’\'{e}tude du dispatching \'{e}conomique,'' \emph{Bulletin de la Soci\'{e}t\'{e} Francaise des \'{e}lectriciens}, vol.~3, pp. 431--447, 1962.

\bibitem{OPF-Lavaei/Duality}
J.~Lavaei and S.~H. Low, ``Zero duality gap in optimal power flow problem,'' \emph{IEEE Transactions on Power Systems}, vol.~27, no.~1, pp. 92--107, 2012.

\bibitem{OPF-Cain/History}
M.~Cain, R.~O'Neill, and A.~Castillo, ``{History of Optimal Power Flow and Formulations},'' \emph{Federal Energy Regulatory Commission Washington DC}, pp. 1--36, 2012.

\bibitem{OPF-Capitanescu/Review}
F.~Capitanescu, ``Critical review of recent advances and further developments needed in ac optimal power flow,'' \emph{Electric Power Systems Research}, vol. 136, pp. 57--68, 2016.

\bibitem{NN-Baker/Warm}
K.~Baker, ``Learning warm-start points for ac optimal power flow,'' in \emph{2019 IEEE 29th International Workshop on Machine Learning for Signal Processing (MLSP)}, 2019, pp. 1--6.

\bibitem{PINN-Dong/PGSim}
W.~Dong, Z.~Xie, G.~Kestor, and D.~Li, ``{Smart-PGSim: Using Neural Network to Accelerate AC-OPF Power Grid Simulation},'' in \emph{SC20: International Conference for High Performance Computing, Networking, Storage and Analysis}, 2020, pp. 1--15.

\bibitem{NN-Deihim/WarmStart}
A.~Deihim, D.~Apostolopoulou, and E.~Alonso, ``Initial estimate of ac optimal power flow with graph neural networks,'' \emph{Electric Power Systems Research}, vol. 234, p. 110782, 2024.

\bibitem{NN-Falconer/Leveraging}
T.~Falconer and L.~Mones, ``{Leveraging Power Grid Topology in Machine Learning Assisted Optimal Power Flow},'' \emph{IEEE Transactions on Power Systems}, vol.~38, no.~3, pp. 2234--2246, 2023.

\bibitem{NN-Fouad/ActiveSet}
F.~Hasan, A.~Kargarian, and J.~Mohammadi, ``Hybrid learning aided inactive constraints filtering algorithm to enhance ac opf solution time,'' \emph{IEEE Transactions on Industry Applications}, vol.~57, no.~2, pp. 1325--1334, 2021.

\bibitem{NN-Zamzam/Learning}
A.~S. Zamzam and K.~Baker, ``Learning optimal solutions for extremely fast ac optimal power flow,'' in \emph{2020 IEEE International Conference on Communications, Control, and Computing Technologies for Smart Grids (SmartGridComm)}, 2020, pp. 1--6.

\bibitem{PINN-Fioretto/Dual}
F.~Fioretto, T.~W. Mak, and P.~Van~Hentenryck, ``Predicting ac optimal power flows: Combining deep learning and lagrangian dual methods,'' in \emph{Proceedings of the AAAI conference on artificial intelligence}, vol.~34, no.~01, 2020, pp. 630--637.

\bibitem{NN-Owerko/GNN}
D.~Owerko, F.~Gama, and A.~Ribeiro, ``Optimal power flow using graph neural networks,'' in \emph{2020 IEEE International Conference on Acoustics, Speech and Signal Processing (ICASSP)}, 2020, pp. 5930--5934.

\bibitem{PINN-Donti/DC3}
P.~L. Donti, D.~Rolnick, and J.~Z. Kolter, ``{DC}3: A learning method for optimization with hard constraints,'' in \emph{International Conference on Learning Representations}, 2021.

\bibitem{PINN-Nellikkath/Physics}
R.~Nellikkath and S.~Chatzivasileiadis, ``{Physics-Informed Neural Networks for AC Optimal Power Flow},'' \emph{Electric Power Systems Research}, vol. 212, p. 108412, 2022.

\bibitem{NN-Singh/Sensitivity}
M.~K. Singh, V.~Kekatos, and G.~B. Giannakis, ``{Learning to Solve the AC-OPF Using Sensitivity-Informed Deep Neural Networks},'' \emph{IEEE Transactions on Power Systems}, vol.~37, no.~4, pp. 2833--2846, 2022.

\bibitem{PINN-Pan/DeepOPF}
X.~Pan, M.~Chen, T.~Zhao, and S.~H. Low, ``Deepopf: A feasibility-optimized deep neural network approach for ac optimal power flow problems,'' \emph{IEEE Systems Journal}, vol.~17, no.~1, pp. 673--683, 2023.

\bibitem{PINN-Liu/GNNDual}
S.~Liu, C.~Wu, and H.~Zhu, ``Topology-aware graph neural networks for learning feasible and adaptive ac-opf solutions,'' \emph{IEEE Transactions on Power Systems}, vol.~38, no.~6, pp. 5660--5670, 2023.

\bibitem{PINN-Owerko/GNN}
D.~Owerko, F.~Gama, and A.~Ribeiro, ``Unsupervised optimal power flow using graph neural networks,'' in \emph{2024 IEEE International Conference on Acoustics, Speech and Signal Processing (ICASSP)}, 2024, pp. 6885--6889.

\bibitem{PINN-Garcia/TypedGNN}
T.~B. Lopez-Garcia and J.~A. Domínguez-Navarro, ``Optimal power flow with physics-informed typed graph neural networks,'' \emph{IEEE Transactions on Power Systems}, vol.~40, no.~1, pp. 381--393, 2025.

\bibitem{PINN-Piloto/CANOS}
\BIBentryALTinterwordspacing
L.~Piloto, S.~Liguori, S.~Madjiheurem, M.~Zgubic, S.~Lovett, H.~Tomlinson, S.~Elster, C.~Apps, and S.~Witherspoon, ``{CANOS: A Fast and Scalable Neural AC-OPF Solver Robust To N-1 Perturbations},'' 2024. [Online]. Available: \url{https://arxiv.org/abs/2403.17660}
\BIBentrySTDinterwordspacing

\bibitem{PINN-Pareek/Bayesian}
\BIBentryALTinterwordspacing
P.~Pareek, A.~Jayakumar, K.~Sundar, D.~Deka, and S.~Misra, ``Optimization proxies using limited labeled data and training time -- a semi-supervised bayesian neural network approach,'' 2025. [Online]. Available: \url{https://arxiv.org/abs/2410.03085}
\BIBentrySTDinterwordspacing

\bibitem{PINN-Park/PrimalDual}
S.~Park and P.~Van~Hentenryck, ``Self-supervised primal-dual learning for constrained optimization,'' \emph{Proceedings of the AAAI Conference on Artificial Intelligence}, vol.~37, no.~4, pp. 4052--4060, 2023.

\bibitem{NN-Rahman2/Augmented}
J.~Rahman, C.~Feng, and J.~Zhang, ``{A learning-augmented approach for AC optimal power flow},'' \emph{International Journal of Electrical Power \& Energy Systems}, vol. 130, p. 106908, 2021.

\bibitem{PINN-Yang/GuidedGNN}
M.~Yang, G.~Qiu, J.~Liu, Y.~Liu, T.~Liu, Z.~Tang, L.~Ding, Y.~Shui, and K.~Liu, ``Topology-transferable physics-guided graph neural network for real-time optimal power flow,'' \emph{IEEE Transactions on Industrial Informatics}, vol.~20, no.~9, pp. 10\,857--10\,872, 2024.

\bibitem{GEN-Jones/OPFLearn}
T.~Joswig-Jones, K.~Baker, and A.~S. Zamzam, ``{OPF-Learn: An Open-Source Framework for Creating Representative AC Optimal Power Flow Datasets},'' in \emph{2022 IEEE Power and Energy Society Innovative Smart Grid Technologies Conference (ISGT)}, 2022, pp. 1--5.

\bibitem{GEN-Ignasi/Rambo}
I.~V. Nadal and S.~Chevalier, ``{Scalable Bilevel Optimization for Generating Maximally Representative OPF Datasets},'' in \emph{2023 IEEE PES Innovative Smart Grid Technologies Europe (ISGT EUROPE)}, 2023, pp. 1--6.

\bibitem{GEN-Gillioz/Entsoe}
M.~Gillioz, G.~Dubuis, and P.~Jacquod, ``A large synthetic dataset for machine learning applications in power transmission grids,'' \emph{Scientific Data}, vol.~12, no.~1, 2025.

\bibitem{GEN-Bugaje/Balancing}
A.-A.~B. Bugaje, J.~L. Cremer, and G.~Strbac, ``Generating quality datasets for real-time security assessment: Balancing historically relevant and rare feasible operating conditions,'' \emph{International Journal of Electrical Power \& Energy Systems}, vol. 154, p. 109427, 2023.

\bibitem{GEN-Venzke/Efficient}
A.~Venzke, D.~K. Molzahn, and S.~Chatzivasileiadis, ``Efficient creation of datasets for data-driven power system applications,'' \emph{Electric Power Systems Research}, vol. 190, p. 106614, 2021.

\bibitem{GEN-Bugaje/Split}
A.-A.~B. Bugaje, J.~L. Cremer, and G.~Strbac, ``Split-based sequential sampling for realtime security assessment,'' \emph{International Journal of Electrical Power \& Energy Systems}, vol. 146, p. 108790, 2023.

\bibitem{GEN-Nellikkath/Enriching}
R.~Nellikkath and S.~Chatzivasileiadis, ``{Enriching Neural Network Training Dataset to Improve Worst-Case Performance Guarantees},'' in \emph{2023 IEEE Belgrade PowerTech}, 2023, pp. 1--6.

\bibitem{GEN-Hu/Worth}
Z.~Hu, H.~Zhang, R.~Yang, Y.~Chen, and H.~Wu, ``Optimal power flow based on physical-model-integrated neural network with worth-learning data generation,'' \emph{CSEE Journal of Power and Energy Systems}, pp. 1--10, 2025.

\bibitem{GEN-Zhang/CSS}
Z.~Zhang, X.~Bai, Z.~Weng, J.~Xiao, and J.~Zhang, ``Generating high-quality datasets with critical state samples for data-driven power system applications,'' \emph{IEEE Access}, vol.~13, pp. 27\,507--27\,515, 2025.

\bibitem{PKG-HEDGeOPF}
\BIBentryALTinterwordspacing
M.~Baù and L.~Perbellini, ``{HEDGeOPF: High-quality, Efficient Dataset Genererator for the AC Optimal Power Flow},'' 2025. [Online]. Available: \url{https://github.com/mttb91/HEDGeOPF.jl}
\BIBentrySTDinterwordspacing

\bibitem{VOL-Vershynin/Probability}
R.~Vershynin, \emph{{High-dimensional probability: An introduction with applications in data science}}.\hskip 1em plus 0.5em minus 0.4em\relax Cambridge University Press, 2018, vol.~47.

\bibitem{VOL-Chalkis/Volesti}
A.~Chalkis and V.~Fisikopoulos, ``{volesti: Volume Approximation and Sampling for Convex Polytopes in R},'' \emph{The R Journal}, vol.~13, no.~2, p. 561, 2021.

\bibitem{PKG-Matpower}
\BIBentryALTinterwordspacing
R.~D. Zimmerman and C.~E. Murillo-Sanchez, ``{MATPOWER User’s Manual, Version 7.1.}'' 2020. [Online]. Available: \url{https://matpower.org/docs/MATPOWER-manual-7.1}
\BIBentrySTDinterwordspacing

\bibitem{PKG-PowerModels}
C.~Coffrin, R.~Bent, K.~Sundar, Y.~Ng, and M.~Lubin, ``Powermodels.jl: An open-source framework for exploring power flow formulations,'' in \emph{Power Systems Computation Conference (PSCC)}, 2018, pp. 1--8.

\bibitem{PKG-JuMP}
M.~Lubin, O.~Dowson, J.~{Dias Garcia}, J.~Huchette, B.~Legat, and J.~P. Vielma, ``{JuMP} 1.0: {R}ecent improvements to a modeling language for mathematical optimization,'' \emph{Mathematical Programming Computation}, vol.~15, p. 581–589, 2023.

\bibitem{SET-Yang/Imbalance}
Y.~Yang, K.~Zha, Y.~Chen, H.~Wang, and D.~Katabi, ``{Delving into Deep Imbalanced Regression},'' in \emph{Proceedings of the 38th International Conference on Machine Learning}, vol. 139, 2021, pp. 11\,842--11\,851.

\bibitem{SET-Gregorius/Simpson}
H.-R. Gregorius and E.~M. Gillet, ``{Generalized Simpson-diversity},'' \emph{Ecological Modelling}, vol. 211, no.~1, pp. 90--96, 2008.

\bibitem{OPF-Aquino/Redundant}
A.~D.~O. Aquino, L.~A. Roald, and D.~K. Molzahn, ``{Identifying Redundant Constraints for AC OPF: The Challenges of Local Solutions, Relaxation Tightness, and Approximation Inaccuracy},'' in \emph{2021 North American Power Symposium (NAPS)}, 2021, pp. 1--6.

\bibitem{GEN-Lovett/OPFData}
\BIBentryALTinterwordspacing
S.~Lovett, M.~Zgubic, S.~Liguori, S.~Madjiheurem, H.~Tomlinson, S.~Elster, C.~Apps, S.~Witherspoon, and L.~Piloto, ``{OPFData: Large-scale datasets for AC optimal power flow with topological perturbations},'' 2024. [Online]. Available: \url{https://arxiv.org/abs/2406.07234}
\BIBentrySTDinterwordspacing

\bibitem{OPF-Babaeine/PGLib}
\BIBentryALTinterwordspacing
S.~Babaeinejadsarookolaee, A.~Birchfield, R.~D. Christie, C.~Coffrin, C.~DeMarco, R.~Diao, M.~Ferris, S.~Fliscounakis, S.~Greene, R.~Huang, C.~Josz, R.~Korab, B.~Lesieutre, J.~Maeght, T.~W.~K. Mak, D.~K. Molzahn, T.~J. Overbye, P.~Panciatici, B.~Park, J.~Snodgrass, A.~Tbaileh, P.~V. Hentenryck, and R.~Zimmerman, ``{The Power Grid Library for Benchmarking AC Optimal Power Flow Algorithms},'' 2021. [Online]. Available: \url{https://arxiv.org/abs/1908.02788}
\BIBentrySTDinterwordspacing

\bibitem{PKG-OPFLearn}
\BIBentryALTinterwordspacing
T.~Joswig-Jones and A.~S. Zamzam, ``{OPFLearn.jl},'' 2023. [Online]. Available: \url{https://github.com/NREL/OPFLearn.jl/releases/tag/v0.1.2}
\BIBentrySTDinterwordspacing

\bibitem{PKG-RAMBO}
\BIBentryALTinterwordspacing
I.~V. Nadal and S.~Chevalier, ``{RAMBO},'' 2023. [Online]. Available: \url{https://github.com/SamChevalier/RAMBO}
\BIBentrySTDinterwordspacing

\bibitem{IPOPT}
A.~W{\"a}chter and L.~T. Biegler, ``On the implementation of an interior-point filter line-search algorithm for large-scale nonlinear programming,'' \emph{Mathematical Programming}, vol. 106, no.~1, pp. 25--57, 2006.

\end{thebibliography}

\end{document}